\documentclass{article}

\usepackage{arxiv}

\usepackage[utf8]{inputenc} % allow utf-8 input
\usepackage[T1]{fontenc}    % use 8-bit T1 fonts
\usepackage{hyperref}       % hyperlinks
\usepackage{url}            % simple URL typesetting
\usepackage{booktabs}       % professional-quality tables
\usepackage{amsfonts}       % blackboard math symbols
\usepackage{nicefrac}       % compact symbols for 1/2, etc.
\usepackage{microtype}      % microtypography
\usepackage{lipsum}
\usepackage{graphicx}

\title{Real-Time Visual Navigation in Huge Image Sets Using Similarity Graphs}

\author{
  Kai Uwe Barthel, Nico Hezel, Konstantin Schall, Klaus Jung\\
  Visual Computing Group, HTW Berlin\\
  Berlin, Germany \\
  \texttt{barthel@htw-berlin.de} \\
}

\begin{document}

\maketitle

\begin{figure}[h]
  %\vspace{27ex} % uncomment this line when using anonymous
  \includegraphics[width=\textwidth]{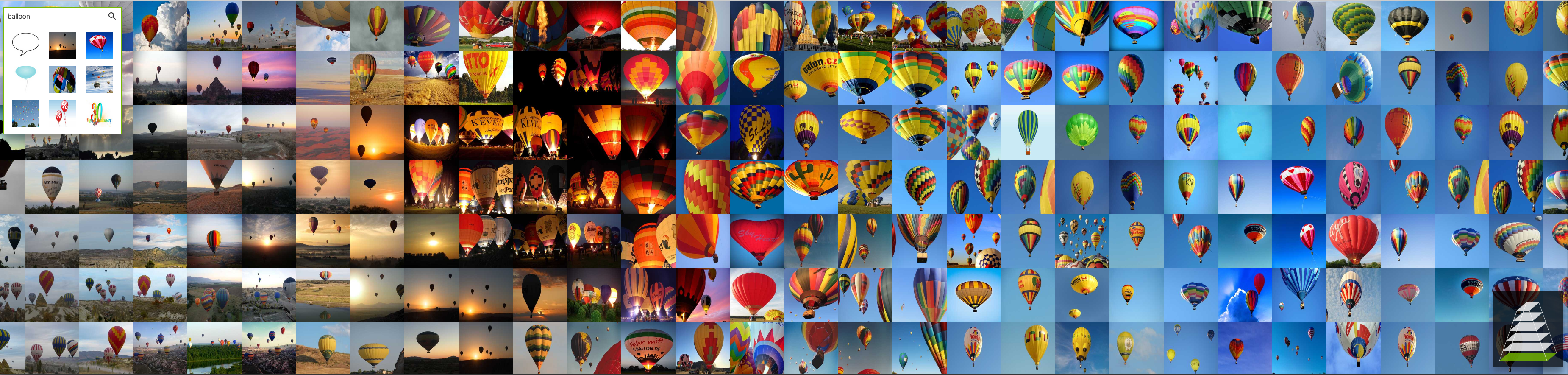}
  \caption{Example visualization of a search with the keyword "balloon". The user may move to related regions of the graph by dragging or zooming the image map or by selecting other "balloon" concepts shown below the search box on the top left.
  }
  \label{fig:teaser}
\end{figure}

\begin{abstract}
Nowadays stock photo agencies often have millions of images. Non-stop viewing of 20 million images at a speed of 10 images per second would take more than three weeks. This demonstrates the impossibility to inspect all images and the difficulty to get an overview of the entire collection. Although there has been a lot of effort to improve visual image search, there is little research and support for visual image exploration. Typically, users start "exploring" an image collection with a keyword search or an example image for a similarity search. Both searches lead to long unstructured lists of result images. 
In earlier publications, we introduced the idea of graph-based image navigation and proposed an efficient algorithm for building hierarchical image similarity graphs for dynamically changing image collections. In this demo we showcase real-time visual exploration of millions of images with a standard web browser. Subsets of images are successively retrieved from the graph and displayed as a visually sorted 2D image map, which can be zoomed and dragged to explore related concepts. Maintaining the positions of previously shown images creates the impression of an "endless map". This approach allows an easy visual image-based navigation, while preserving the complex image relationships of the graph.    
\end{abstract}

% keywords can be removed

\section{Introduction}
In archives, images are typically represented by their keywords and/or high-dimensional visual feature vectors. 
If too many pictures are displayed at the same time, people quickly lose the overview. Sorting images by their visual similarity can help to view and recognize more images simultaneously. Conventional dimension reduction techniques which project feature vectors to two dimensions, cannot be used since their projections produce unequally distributed and overlapping images. 
If images are to be sorted on a dense regular grid (map), self-organizing maps (SOM) \cite{Kohonen:1997:SM:261082} or self-sorting maps (SSM) \cite{Strong:2014:SME:2719262.2719635} can be used. When dealing with very large collections, hierarchical visualization and navigation techniques are necessary. Picsbuffet \href{https://www.picsbuffet.com}{(www.picsbuffet.com)} is an example of such a scheme using 1 million stock photos from Pixabay. 

However, the similarity relationships between images are very complex and cannot be preserved with 2D projections.
Another important disadvantage of static regular 2D maps is their inability to handle changes of the image collection. Removed images will result in holes in the image map, whereas for newly added images a new sorting process is unavoidable. Graph-based approaches can handle these changes and are better able to represent the complex image relationships.

In this demo we present a graph-based system for visually exploring and navigating very large, continuously changing image sets. The similarities of the image feature vectors are used to build a hierarchical image graph. When evaluating possible navigation and visualization options, we found that the best user experience is achieved by dynamically projecting sub-graphs onto a visually sorted 2D image map. Dragging this map and zooming (changing the hierarchy level/layer) allows the user to explore the image collection similar to an interactive map service such as \textit{Google Maps}.

\section{Proposed Method / Building Blocks}

\subsection{Generation of Feature Vectors}
To build the image graph we use semantic feature vectors generated from a ResNet50 \cite{He15} which is retrained for the specific task of image retrieval using the \textit{Google Landmarks} \cite{Noh17} and \textit{ImageNet} \cite{Russakovsky14} data sets with the newly proposed \textit{Nonlinear Rank Approximation} loss function \cite{NRA19}. Furthermore, we perform aggregation of regional convolutional activations similar to \cite{Tolias16} which is done during training in an end-to-end manner. Images of 299 pixels on both sides are used for fast feature extraction. The obtained 2048-dimensional feature vectors achieve a \textit{mean average precision score} of 93.5 on the Holidays data set \cite{Holidays}. These feature vectors are further compressed to 64 dimensions and cast to bytes for efficient similarity search and storage. A second handcrafted low-level feature vector describes the visual appearance of the images.   

\subsection{Building and Manipulating the Graph}
The image graph has to be constructed in such a way that similar images are connected and related images can be reached by navigating the edges of the graph. The \textit{graph quality} is defined as the average similarity of all connected image nodes. 

In \cite{Barthel:2017:VBM:3078971.3079016} we propose an algorithm for improving quartic similarity graphs. Starting from an initial random quartic graph, an iterative process applies sequences of reconnections to improve the graph quality. It uses the similarity of the semantic feature vectors to connect related images. In \cite{Hezel:2018:DCM:3206025.3206093} we extend this approach to dynamically create and manipulate a graph. This allows adding and removing nodes (images) at any time without rebuilding the graph from scratch. In addition, hierarchical graph layers are build dynamically which help to summarize lower layers and ensure that the overview is not lost when navigating the graph. Combining the two approaches allows generating hierarchical image graphs, which can grow and improve over time.

\subsection{Fast Visual Image Sorting}
For displaying related images, we retrieve images from sub-graphs (see next section for details). To generate a visual pleasing image arrangement we use both feature vector types and sort them with an improved version of a Self Sorting Map (SSM). 
%Given a set of images and a similarity measure between each pair of them, the SSM places each image into a unique cell of a structured layout, where the most similar images are placed together and the unrelated ones are spread apart. 
In \cite{doi:10.1002/9781119376996.ch11} we show how to improve the sorting quality by combining the advantages of SOM and SSM. This approach allows to sort hundreds of images (feature vectors) in fractions of a second. 

\subsection{Visual Navigation}
For a keyword search we determine regions of the graph with suiting images. From the most prominent region we select a representative image. The edges of this image node are recursively followed until a desired number of neighbor images has been retrieved. These images are visually sorted and displayed (see Figure \ref{fig:teaser}). If the user cannot find the desired image, the map can be dragged to the region where the searched images are expected. Dragging moves some images out of view and leaves empty space on the opposite side. The inverse direction of the mouse drag indicates images of interest (at the border to the empty region). Following the graph-edges of these images, we retrieve new images and place them into the empty region of the map. This again is done by visual sorting, however previously displayed images do not change their positions. To emphasize a realistic navigation, image positions are cached, so the user will find previously seen images at the same positions if dragging back to regions visited before.

At any time the user may also choose any image as new center image for which a new retrieval is started in the way as described above. Zooming out shows an overview of related concepts, zooming in will show more similar images. In these cases the closest sub-graph of the next hierarchy level is selected, visually sorted and displayed. A smooth transition from the old map to the new map is performed to avoid the user getting lost.

\section{Implementation}
For an efficient web browser implementation we heavily utilize a client server architecture. The image feature vectors are extracted on the server side. The extension, improvement and storage of the graph is performed continuously in a separate process. 

The memory requirement per image consists of the image ID, the four IDs of the connected images plus the two feature vectors (semantic: 64 bytes \& low-level visual: 50 bytes). In total only 134 bytes per image are needed to describe the graph. A hierarchical graph of 100 million images can be stored with less than 20 GB.  

For each drag or search operation of a user the client sends a request of the new position or a keyword search string, as well as additional view port information. The server filters and re-sorts the images and sends the new 2D map with image IDs and their positions. The client constructs image URLs from the received IDs and downloads the corresponding images. The images are displayed using a single HTML canvas to benefit from GPU accelerated drawing. Since modern browsers cache and download multimedia data very efficiently, managing of the image data is entrusted to the browser if working with URLs. Therefore this architecture utilizes the browser as a pure view layer and performs the graph-based calculations on the server. 

A video including screencasts of the visual image browsing scheme and a demo website can be found at:

\noindent \href{https://www.visual-computing.com/project/graph/}{www.visual-computing.com/project/graph}

\bibliographystyle{unsrt}  
\bibliography{references}  %%% Remove comment to use the external .bib file (using bibtex).

\end{document}